# On the solution of the Heaviside - Klein - Gordon thermal equation for heat transport in graphene

Magdalena Pelc

Institute of Physics, Maria Curie - Sklodowska University, Lublin, Poland

Abstract

We report studies of the solution of the Heaviside - Klein - Gordon thermal equation. As the result it is shown that the solution consists of two components: the fast thermal wave and slow diffusion for very large (compared to relaxation time) time period. We argue that the fast thermal wave can be recognized as the indication of the ballistic heat transport. As an example we consider the ballistic heat transport in graphene.

## 1. Introduction

In the desription of the evolution of any physical system, it is mandatory to evaluate as accurately as possible the order of magnitude of different characteristic time scales, since their relationship with the time scale of observation (the time during which we assume our description of the system is valid) will determine along with the relevant equation pattern.

The advent of attosecond laser pulses opens the new field of investigation of the quantum phenomena. As all measured relaxation times are much longer the attosecond laser pulses "observe" the generic quantum nature of the phenomena, not averaged over time.

In monograph [2] the theoretical framework for transport processes generated by attosecond laser pulses was formulated. It was shown that the master equation is the hyperbolic transport equation:

$$\frac{1}{v^2}\frac{\partial^2 T(\vec{r},t)}{\partial t^2} - \Delta T(\vec{r},t) + qT(\vec{r},t) = 0 \qquad (1.1)$$

$$q = \frac{2Vm}{\hbar^2} - \left(\frac{mv}{2\hbar}\right)^2 \qquad (1.2)$$

$$v = \alpha c,$$

where $\alpha = 1/137$ is the electromagnetic coupling constant and $c$ is the light velocity, $m$ - heat carrier mass.

Depending on the sign of the $q$ the equation is the Heaviside equation ($q < 0$) or Klein - Gordon equation ($q > 0$). For $q = 0$ Eq.(1.1) is the wave equation which describes the ballistic, quasi-free propagation of carriers.

Recent measurement of the electrical and thermal properties of graphene [1] shed new light on the application of the equation (1.1) to the investigation of the transport phenomena.

As an example we invoke the ballistic transport in graphene which as can be seen from Eq.(1.1) is the result of the $q = 0$, moreover it occurs that the velocity of fermions in graphene is of the order of $10^6$ m/s which agrees with $v = \alpha c$.

The aim of this paper is the solution of the equation (1.1) for the Klein - Gordon and Heaviside branches.

## 2. The model equations

We will describe the thermal energy transport in graphene [1] within the theory of the hyperbolic transport equation [2]. Given the initial value problem for hyperbolic thermal equation

$$u_{tt} - \gamma^2 u_{xx} + c^2 u = F(x,t) \tag{2.1}$$

$$-\infty < x < \infty, \quad t > 0$$

with the initial conditions

$$u(x,0) = f(x), \qquad u_t(x,0) = g(x), \tag{2.2}$$

$$-\infty < x < \infty,$$

the solution at an arbitrary point $(\xi,\tau)$ is given by [1]

$$u(\xi,\tau) = \int_0^T \int_{-\infty}^\infty FK\,dx\,dt + \int_{-\infty}^\infty \left[ gK(x,0;\xi,\tau) - f\frac{\partial K}{\partial t}(x,0;\xi,\tau) \right] dx, \tag{2.3}$$

where $\tau < T$ and $K(x,0;\xi,\tau)$ is the solution of the equation

$$\frac{\partial^2 K}{\partial t^2} - \gamma^2 \frac{\partial^2 K}{\partial x^2} + c^2 K = \delta(x-\xi)\delta(t-\tau). \tag{2.4}$$

As can be seen from equation (2.4) $K(x,0;\xi,\tau)$ is the Green function for the equation (2.1) and fulfils the formula

$$K(x,t;\xi,\tau) = \frac{1}{2\gamma} J_0\left[ \frac{c}{\gamma} \sqrt{\gamma^2(t-\tau)^2 - (x-\xi)^2} \right] \quad \text{for } |x-\xi| < \gamma(t-\tau)$$

$$K(x,t;\xi,\tau) = 0 \quad \text{for } |x-\xi| > \gamma(t-\tau) \tag{2.5}$$

and $J_0(y)$ is the Bessel function of the order 0.

Introducing the Heaviside step function $H(z)$ we express formulae (2.5) as

$$K(x,t;\xi,\tau) = \frac{1}{2\gamma} J_0\left[\frac{c}{\gamma}\sqrt{\gamma^2(t-\tau)^2 - (x-\xi)^2}\right] * H[x - \gamma t - (\xi - \gamma\tau)]H[\xi + \gamma\tau - (x + \gamma t)]$$

(2.6)

Since $J_0(0) = 1$ we see that (2.6) reduces to the Green function for the one dimensional wave equation if we set $c = 0$.

We now show that $K(x,0;\xi,\tau)$ satisfies (2.4). To that aim we set

$$K = J_0 \hat{K},$$ (2.7)

where $\hat{K}$ is the Green function for the wave equation and satisfies

$$\frac{\partial^2 \hat{K}}{\partial t^2} - c^2 \frac{\partial^2 \hat{K}}{\partial x^2} = \delta(x-\xi)\delta(t-\tau)$$ (2.8)

with $c$ replaced by $\gamma$. Then

$$\frac{\partial^2 K}{\partial t^2} - \gamma^2 \frac{\partial^2 K}{\partial x^2} + c^2 K = \left\{\frac{\partial^2 J_0}{\partial t^2} - \gamma^2 \frac{\partial^2 J_0}{\partial x^2} + c^2 J_0\right\}\hat{K}$$

$$+ 2\left\{\frac{\partial J_0}{\partial t}\frac{\partial \hat{K}}{\partial t} - \gamma^2 \frac{\partial J_0}{\partial x}\frac{\partial \hat{K}}{\partial x}\right\}$$ (2.9)

$$+ \left\{\frac{\partial^2 \hat{K}}{\partial t^2} - \gamma^2 \frac{\partial^2 \hat{K}}{\partial x^2}\right\}J_0$$

For $J_0\left[\frac{c}{\gamma}\sqrt{\gamma^2(t-\tau)^2 - (x-\xi)^2}\right]$ the first term in (2.9) is equal zero. Also we have

$$2\left\{\frac{\partial J_0}{\partial t}\frac{\partial \hat{K}}{\partial t} - \gamma^2 \frac{\partial J_0}{\partial x}\frac{\partial \hat{K}}{\partial x}\right\} = \frac{cJ_0'\left[\frac{c}{\gamma}\right]\sqrt{\gamma^2(t-\tau)^2 - (x-\xi)^2}}{\sqrt{\gamma^2(t-\tau)^2 - (x-\xi)^2}} *$$

$$* \left\{\begin{array}{l}[-\gamma(t-\tau) + (x-\xi)]\delta[x-\gamma t - (\xi - \gamma t)]H[\xi + \gamma\tau - (x - \gamma t)] + \\ [-\gamma(t-\tau) - (x-\xi)]H[x - \gamma t - (\xi - \gamma t)]\delta[\xi + \gamma\tau - (x + \gamma t)]\end{array}\right\}.$$ (2.10)

The expression (2.100 vanishes since

$$[-\gamma(t-\tau) + (x-\xi)]\delta[x - \gamma t - (\xi - \gamma t)] =$$
$$[-\gamma(t-\tau) + (x-\xi)]\delta[-\gamma(t-\tau) + (x-\xi)] = 0.$$ (2.11)

$$[-\gamma(t-\tau) - (x-\xi)]\delta[\xi + \gamma\tau - (x + \gamma t)] =$$
$$[-\gamma(t-\tau) - (x-\xi)]\delta[-\gamma(t-\tau) - (x-\xi)] = 0.$$ (2.12)

on using $x\delta(x) = 0$. Finally we have

$$\left\{\frac{\partial^2 \hat{K}}{\partial t^2} - \gamma^2 \frac{\partial^2 \hat{K}}{\partial x^2}\right\} J_0 = \delta(x-\xi)\delta(t-\tau) J_0\left[\frac{c}{\gamma}\sqrt{\gamma^2(t-\tau)^2 - (x-\xi)^2}\right] =$$
$$= \delta(x-\xi)\delta(t-\tau) J_0(0) = \delta(x-\xi)\delta(t-\tau). \tag{2.13}$$

It may be noted that with $c = i\hat{c}$ (where $i = \sqrt{-1}$) in the equation (2.1) we obtain Heaviside equation

$$\frac{\partial^2 K}{\partial t^2} - \gamma^2 \frac{\partial^2 K}{\partial x^2} - \hat{c}^2 K = \delta(x-\xi)\delta(t-\tau). \tag{2.14}$$

Since $J_0(ix) = I_0(x)$, the modified Bessel function of zero order we obtain for $K$ in place (2.6)

$$K(x,t;\xi,\tau) = \frac{1}{2\gamma} I_0\left[\frac{\hat{c}}{\gamma}\sqrt{\gamma^2(t-\tau)^2 - (x-\xi)^2}\right] * H[x - \gamma t - (\xi - \gamma\tau)] H[\xi + \gamma\tau - (x+\gamma t)] \tag{2.15}$$

is the Green function.

In the double integral (2.3) we have $K = 0$ for $t > \tau$ so that the limit in the $t$ integral extends only up to $\tau$. Also from (2.5) we conclude that $K$ vanishes unless $|x - \xi| < \gamma(\tau - t)$ and this is equivalent to

$$\xi - \gamma(\tau - t) < x < \xi + \gamma(\tau - t). \tag{2.16}$$

Therefore we obtain

$$\int_0^T \int_{-\infty}^{\infty} FK dx dt = \frac{1}{2\gamma} \int_0^T \int_{\xi-\gamma(\tau-t)}^{\xi+\gamma(\tau-t)} F(x,t) J_0\left[\frac{c}{\gamma}\sqrt{\gamma^2(t-\tau)^2 - (x-\xi)^2}\right] dx dt. \tag{2.17}$$

Further we have

$$K(x,0;\xi,\tau) = \frac{1}{2\gamma} J_0\left[\frac{c}{\gamma}\sqrt{\gamma^2\tau^2 - (x-\xi)^2}\right] * H[x - (\xi - \gamma\tau)] H[\xi + \gamma\tau - x] \tag{2.18}$$

so that

$$\int_{-\infty}^{\infty} g(x) K(x,0;\xi,\tau) dx = \frac{1}{2\gamma} \int_{\xi-\gamma\tau}^{\xi+\gamma\tau} g(x) J_0\left[\frac{c}{\gamma}\sqrt{\gamma^2\tau^2 - (x-\xi)^2}\right] dx. \tag{2.19}$$

Since the product of the Heaviside function vanishes outside the interval $(\xi - \gamma\tau, \xi + \gamma\tau)$. Finally

$$\partial K(x,0;\xi,\tau) = -\frac{c\tau}{2}\frac{J_0'\left[\frac{c}{\gamma}\sqrt{\gamma^2\tau^2-(x-\xi)^2}\right]}{\sqrt{\gamma^2\tau^2-(x-\xi)^2}}H[x-(\xi-\gamma\tau)]H[\xi+\gamma\tau-x]$$

$$-\frac{1}{2}J_0\left[\frac{c}{\gamma}\sqrt{\gamma^2\tau^2-(x-\xi)^2}\right]\delta[x-(\xi-\gamma\tau)]H[\xi+\gamma\tau-x] \quad (2.20)$$

$$-\frac{1}{2}J_0\left[\frac{c}{\gamma}\sqrt{\gamma^2\tau^2-(x-\xi)^2}\right]H[x-(\xi-\gamma\tau)]\delta[\xi+\gamma\tau-x].$$

In view of the substitution property of the delta function, the last two terms in (2.20) reduce to $-\tfrac{1}{2}\delta[x-(\xi-\gamma\tau)]-\tfrac{1}{2}\delta[x-(\xi+\gamma\tau)]$ since $J_0(0)=1$ and $H(2\gamma\tau)=1$.

Therefore we obtain

$$\int_{-\infty}^{\infty} f(x)\frac{\partial K}{\partial t}(x,0;\xi,\tau)dx = -\frac{c\tau}{2}\int_{\xi-\gamma\tau}^{\xi+\gamma\tau} f(x)\frac{J_0'\left[\frac{c}{\gamma}\sqrt{\gamma^2\tau^2-(x-\xi)^2}\right]}{\sqrt{\gamma^2\tau^2-(x-\xi)^2}}dx \quad (2.21)$$

$$-\frac{1}{2}f(\xi-\gamma\tau)-\frac{1}{2}f(\xi+\gamma\tau)$$

Combining these results and noting that $-J_0'(x)=J_1(x)$ the Bessel function of order one, gives the solution $u(x,t)$ of the initial value problem (2.2) as

$$u(x,t) = \frac{f(x-\gamma t)+f(x+\gamma t)}{2} + \frac{1}{2\gamma}\int_{x-\gamma t}^{x+\gamma t} g(\xi)J_0\left[\frac{c}{\gamma}\sqrt{\gamma^2\tau^2-(x-\xi)^2}\right]d\xi$$

$$-\frac{c\tau}{2}\int_{x-\gamma t}^{x+\gamma t} f(\xi)\frac{J_1\left[\frac{c}{\gamma}\sqrt{\gamma^2\tau^2-(x-\xi)^2}\right]}{\sqrt{\gamma^2\tau^2-(x-\xi)^2}}d\xi \quad (2.22)$$

$$+\frac{1}{2\gamma}\int_0^t \int_{x-\gamma(t-\tau)}^{x+\gamma(t-\tau)} F(\xi,\tau)J_0\left[\frac{c}{\gamma}\sqrt{\gamma^2(t-\tau)^2-(x-\xi)^2}\right]d\xi d\tau.$$

This solution formula reduces to that for the Cauchy problem for the inhomogenous wave equation if we set $c=0$. Also if we set $c=i\hat{c}$ in (2.22) and note that $J_0(iz)=I_0(z)$ and $J_1(iz)=iI_1(z)$ we obtain as the solution Heaviside equation

$$u_{tt}-\gamma^2 u_{xx}+\hat{c}^2 u = F(x,t) \quad (2.22)$$

$$-\infty < x < \infty, \quad t>0$$

with the initial condition (2.3)

$$u(x,t) = \frac{f(x-\gamma t)+f(x+\gamma t)}{2} + \frac{1}{2\gamma}\int_{x-\gamma t}^{x+\gamma t} g(\xi)I_0\left[\frac{\hat{c}}{\gamma}\sqrt{\gamma^2 t^2 - (x-\xi)^2}\right]d\xi$$

$$+ \frac{\hat{c}t}{2}\int_{x-\gamma t}^{x+\gamma t} f(\xi) \frac{I_1\left[\frac{\hat{c}}{\gamma}\sqrt{\gamma^2 t^2 - (x-\xi)^2}\right]}{\sqrt{\gamma^2 t^2 - (x-\xi)^2}} d\xi \qquad (2.24)$$

$$+ \frac{1}{2\gamma}\int_0^t \int_{x-\gamma(t-\tau)}^{x+\gamma(t-\tau)} F(\xi,\tau) I_0\left[\frac{\hat{c}}{\gamma}\sqrt{\gamma^2(t-\tau)^2 - (x-\xi)^2}\right] d\xi d\tau.$$

## 3. Ballistic heat transport in graphene

Very important reason for the interest in graphene is a unique nature of its charge carriers. In condensed matter physics the Schrödinger equation rules the world, usually being quite sufficient to describe electronic properties of materials. Graphene is an exception: its charge carriers mimic relativistic particles and are easier and more natural to describe starting with the relativistic equations: Klein - Gordon and Dirac rather than the Schrödinger equation. although there is nothing particularly relativistic about electrons moving around carbon atoms, their interaction with a periodic potential of graphene lattice gives rise to new quasiparticles that at low energies are accurately described by (2+1) dimensional equation with an effective speed of light $v \approx 10^6$ m/s. These quasiparticles, called massless Dirac fermions, can be seen as electrons that lost their rest mass $m_0$ or as neutrinos that acquired the electron charge.

In the monograph [2] the quantum thermal equation for the heat transport was formulated.

$$\frac{1}{v^2}\frac{\partial^2 T}{\partial t^2} + \frac{m}{\hbar}\frac{\partial T}{\partial t} + \frac{2Vm}{\hbar^2}T - \frac{\partial^2 T}{\partial x^2} = F(x,t). \qquad (3.1)$$

The solution of equation (3.1) can be written as

$$T(x,t) = e^{\frac{1}{2}\tau} u(x,t). \qquad (3.2)$$

After substituting formula (3.2) into (3.1) we obtain new equation

$$\frac{1}{v^2}\frac{\partial^2 u}{\partial t^2} - \frac{\partial^2 u}{\partial x^2} + qu(x,t) = e^{\frac{1}{2}\tau} F(x,t), \qquad (3.3)$$

and

$$q = \frac{2Vm}{\hbar^2} - \left(\frac{m\upsilon}{2\hbar}\right)^2. \tag{3.4}$$

Equation (3.3) can be written as

$$\frac{\partial^2 u}{\partial t^2} - \upsilon^2 \frac{\partial^2 u}{\partial x^2} + q\upsilon^2 u(x,t) = G(x,t), \tag{3.5}$$

where

$$G(x,t) = \upsilon^2 e^{t/2\tau} F(x,t).$$

Equation (3.5) has the same form as the equation (1.1). As can be seen from Eq.(3.5) for $q = 0$ we obtain ballistic heat transport with velocity $\upsilon = \alpha c \approx 10^6$ m/s as in graphene.